\documentclass{svproc}

\usepackage{url}

\usepackage{tabularx}
\usepackage{multirow}
\usepackage{color}
\usepackage{booktabs}
\usepackage[square,comma,numbers,sort&compress,sectionbib]{natbib}
\usepackage{enumerate}
\usepackage{enumitem}
\usepackage{graphicx}
\usepackage{cleveref}
\usepackage{makecell}

\begin{document}
\mainmatter

\title{CIR at iKAT SCAI 2026: Exploring Clarification Need Prediction in Agentic Conversational Search}
\titlerunning{CIR@ iKAT SCAI 2026}

\author{Nolwenn Bernard$^{\dagger}$ \and Jüri Keller$^{\dagger}$ \and Philipp Schaer$^{\dagger}$}
\authorrunning{Bernard et al.}

\tocauthor{Nolwenn Bernard, Jüri Keller, Philipp Schaer}

\institute{$^{\dagger}$TH Köln, Cologne, Germany\\
\email{$^{\dagger}$\{nolwenn.bernard, jueri.keller, philipp.schaer\}@th-koeln.de}}

\maketitle

\begin{abstract}
    This paper presents the participation of the Cologne Information Retrieval group in the iKAT SCAI 2026 shared task. We use an agentic conversational search system, equipped with tools for query rewriting, retrieval and reranking, answer generation, and clarification need prediction and clarification question generation. We experiment with two different neural clarification need prediction models.
\end{abstract}

\section{Introduction}
\label{sec:intro}

A shift towards conversational search has been observed in recent years, with the emergence of Large Languages Models~(LLMs) and their integration into the search process. This has resulted in a growing interest in the development of conversational search systems that are capable of engaging in natural language interactions with users to support their information needs. The \emph{Interactive Knowledge Assistance Track}~(iKAT) shared task, and its predecessor \emph{Conversational Assitant Track}~(CAsT)~\citep{Dalton:2020:arXiv}, provide an environment for the development and evaluation of conversational search systems. The iKAT shared task specifically focuses on the development of systems capable of providing personalized, context-aware, and informative responses to users. The conversational systems are evaluated interactively with simulated users conditioned on a Personal Textual Knowledge Base~(PTKB) to test the robustness to a diverse user population. The PTKB comprises a set of statements related to the personality, preferences, and interests.

Our participation focuses on the mixed-initiative ability of the conversational system, particularly for clarification need prediction. Rather than adopting a conventional pipeline, our system employs a naive agentic architecture. 
The different components of the traditional pipeline, i.e., query rewriting, retrieval, re-ranking, and response generation, and the mixed-initiative ones are considered as tools that can be called by the orchestrator agent. For the clarification need prediction, we experiments with two neural classification models: MuSIc~\citep{Meng:2023:CIKM} and Zef-CNP BERT-based model~\citep{Lu:2025:ECIR}.

Empirical results indicate marginal differences between the two submitted runs, with a slight advantage observed when employing the Zef-CNP  BERT-based model. The dialogue-level evaluation indicates moderate mixed-initiative performance, averaging approximately 3 out of 5 across user profiles. Conversely, passage ranking metrics reveals retrieval as a critical area for improvement, with low retrieval rates.

\section{Related Work}
\label{sec:related_work}

Multiple requirements have been identified for conversational search systems, including being able to understand user intents, provide context-aware and personalized responses, and support mixed-initiative interactions~\citep{Zamani:2023:FnTIR}. 
In order to satisfy these requirements, multiple architectures have been proposed over the years. We identify two categories of architectures: (1) pipeline-based architectures and (2) agentic architectures. Pipeline-based architectures comprise different components that are executed based on a predefined workflow. The typical components are query rewriting, retrieval, re-ranking, response generation, and in some cases mixed-initiative components.
This type of architecture remains the most common in both iKAT and CAsT shared tasks~\citep{Owoicho:2022:TREC,Aliannejadi:2025:TREC}.
While agentic architectures also comprise different components, they are orchestrated by an agent that decides which component to call and in which order based on the current state of the conversation. This type of architecture has been receiving more attention with the emergence of generative AI~\citep{Meng:2026:WWW}. Recent work on agentic conversational search systems uses multiple rounds of reasoning and search engine calls to generate responses~\citep{Mo:2026:ACL,Lupart:2026:ACL}. After the reception of a user utterance, the agent produces a trajectory including the reasoning, the search queries issued to a search engine, the retrieved documents, and the final response. The agent is trained to produce trajectories that maximize the quality of the final response.
Our conversational search system follows an agentic architecture, however, the reasoning is completely performed by the orchestrator without any additional training and the tools correspond to the components of a traditional pipeline-based architecture. 
 
For our runs, we focus on the mixed-initiative ability of the conversational search system. Mixed-initiative interactions refers to the ability of each participant in a conversation to take the initiative and contribute to the conversation~\citep{Allen:1999:IEEE}. In the context of conversational search, mixed-initiative interactions often relate to when to ask and how to formulate clarification questions~\citep{Zamani:2023:FnTIR}. 
Identifying when the system should take ask a clarification question is often viewed as a (binary) classification problem, e.g,~\citep{Xu:2019:EMNLP,Lu:2025:ECIR,Wang:2021:WWW}. For our submissions, we first consider the MuSIc model~\citep{Meng:2023:CIKM}, which encodes conversational context and predicts the sequence of system initiative or non-initiative actions for utterances in the conversation and the next one. This model can be trained to support multiple mixed-initiative actions (e.g., repeat question, ask clarification question, ask follow-up question) depending on the training data. Second, we have a model trained using the Zef-CNP framework~\citep{Lu:2025:ECIR}. The framework generates a synthetic dataset using LLMs to then train a classification model for clarification need prediction, which is finally used for inference. The Zef-CNP framework is designed to be used with any classification model, following \citet{Lu:2025:ECIR}, we fine-tune a BERT model for our submission.
Formulating a clarification question can be done in multiple ways. Previous work has explored selecting a clarification question from a pre-defined pool~\citep{Sekulic:2024:ECIR,Aliannejadi:2019:SIGIR}, template-based generation~\citep{Zamani:2020:WWW}, and generative approaches~\citep{Sekulic:2021:ICTIR,Zamani:2020:WWW,Wang:2023:WWW}. In our work, we use a generative approach to produce clarification questions in a zero-shot manner.

\section{Approach}
\label{sec:approach}

\begin{figure}
    \includegraphics[width=\textwidth,keepaspectratio]{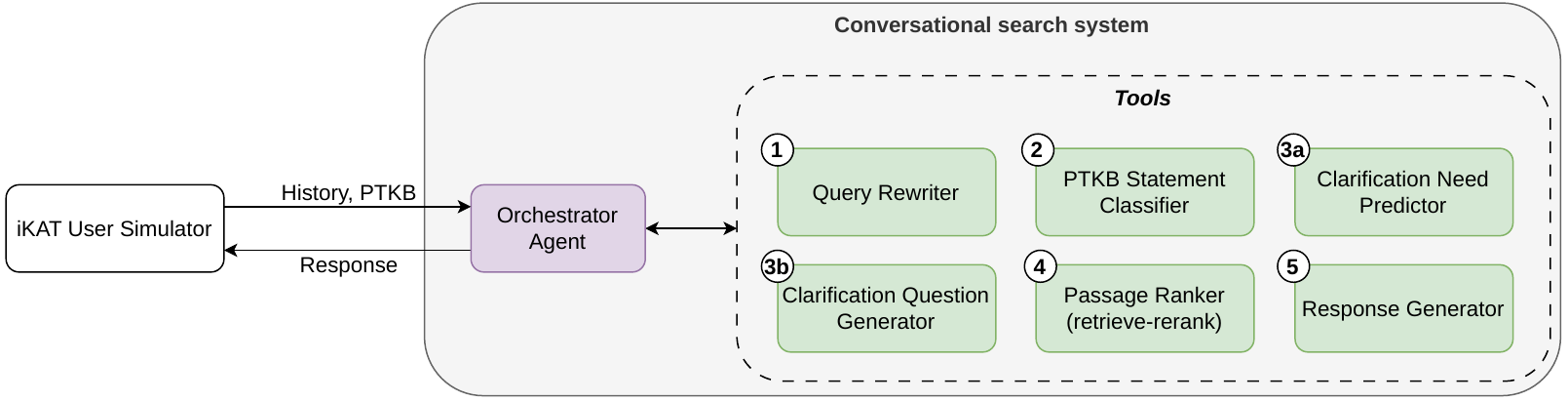}
    \caption{Overview of our conversational search system, including the orchestrator and tools. The tools are numbered in the order they should be called by the orchestrator, although the linearity of the pipeline is not strictly enforced.}
    \label{fig:system_overview}
\end{figure}

\begin{figure}
    \begin{center}
        \ttfamily
        \fbox{\begin{minipage}{\textwidth}
            You are a conversational search assistant following a structured pipeline. For each user utterance, reason step by step:\\
            1. Call `rewrite\_query` to resolve coreferences and produce a standalone search query. \\
            2. Call `select\_ptkb\_statements` to choose relevant personal background statements (if any) to include in the query context. \\
            3. Call `needs\_clarification` to check if the query is ambiguous or too vague. \\
                - If True: call `generate\_clarification\_question` and return that question as your response. \\
                - If False: proceed with the retrieval pipeline below. \\
            4. Call `retrieve\_and\_rerank` with the rewritten query to fetch and rerank candidate passages. \\
            5. Call `generate\_answer` with the rewritten query, retrieved passages, and selected personal background statements to produce the final grounded response. This is the response you return to the user. \\
            CRITICAL RULES: \\
            - Never answer from your own knowledge. You must always use the tools in the pipeline. \\
            - The pipeline is strictly linear. You cannot call tools from previous steps once you have moved on to the next step. \\
            - The response to the user cannot exceed 250 tokens. If your answer is too long, you must shorten it to fit the limit. \\
        \end{minipage}}
    \end{center}
    \caption{Prompt provided to the orchestrator for decision-making.}
    \label{fig:orchestrator_prompt}
\end{figure}

We briefly describe the different components of our conversational search system, an overview is shown in Figure~\ref{fig:system_overview}. 

\paragraph{Query Rewriting.} We use a T5-based model fine-tuned on the CANARD dataset for query rewriting (\texttt{castorini/t5-base-canard}\footnote{\url{https://huggingface.co/castorini/t5-base-canard}}). This model is well established for the task of query rewriting in conversational search, and is publicly available on the Hugging Face Transformers library.\footnote{\url{https://github.com/huggingface/transformers}}. The rewriter takes the history of the conversation including the latest user utterance as input and generates a rewritten query that aims to resolve coreferences and produce a standalone search query.

\paragraph{PTKB Statement Classification.} We classify the relevant personal statements from the PTKB in a zero-shot manner using the model \texttt{google/gemma-4-26B-A4B-it}.\footnote{\url{https://recipes.vllm.ai/Google/gemma-4-26B-A4B-it}} The model is prompted to provide a list of relevant personal statements given the history including the latest user utterance and all possible PTKB statements. 

\paragraph{Passage Ranking.} We use a traditional two-stage pipeline for passage ranking, consisting of a retrieval step followed by a re-ranking step. For the first pass, we use BM25 ($k1=0.9, b=0.4$) to retrieve the top 1000 passages given the rewritten query from the Pyserini\footnote{\url{http://pyserini.io/}} index provided by the organizers. These passages are then re-ranked using the default monoT5 re-ranker from PyTerrier\_t5\footnote{\url{https://github.com/terrierteam/pyterrier_t5}} (\texttt{castorini/monot5-base-msmarco}\footnote{\url{https://huggingface.co/castorini/monot5-base-msmarco}}).

\paragraph{Clarification Need Prediction.} We identify whether a clarification question should be asked, i.e., this task is formulated as a binary classification problem. We support two existing models. 
First, we have the MuSIc model~\citep{Meng:2023:CIKM} trained on MSDialog~\citep{Qu:2018:SIGIR} and fine-tuned on ClariQ~\citep{Aliannejadi:2021:EMNLP}. This model takes the history of the conversation including the latest user utterance as input.
Second, we have a BERT-based model trained for clarification need prediction using the Zef-CNP framework~\citep{Lu:2025:ECIR}. Unlike the MuSIc model, this model takes the rewritten query as input.

\paragraph{Clarification Question Generation.} We generate a clarification question in a zero-shot manner upon the prediction of a clarification need. We use the model \texttt{google/gemma-4-26B-A4B-it} to generate a clarification question given the history of the conversation including the latest user utterance and the relevant personal statements from the PTKB. 

\paragraph{Response Generation.} We generate a response in a zero-shot manner using the model \texttt{google/gemma-4-26B-A4B-it}. The model is provided the retrieved passages, the relevant personal statements from the PTKB, and the history of the conversation including the last user utterance. 

\paragraph{Orchestrator.} The orchestrator is responsible for calling the appropriate tools based on its high-level instruction and the conversation history. A description of the tools, the traditional rewrite-retrieval-rerank-answer pipeline, and instructions are provided to the orchestrator to guide its decision-making process; the prompt is shown in Figure~\ref{fig:orchestrator_prompt}. We note that linearity of the pipeline described in the instructions prompt is not strictly enforced, that is, the orchestrator decides which tools to call and in which order (including the possibility of calling the same tool multiple times). 
For the orchestrator, we use the model \texttt{google/gemma-4-26B-A4B-it}.

\subsection{Implementation Details}

The backbone of our conversational search system relies on the Pydantic AI framework\footnote{\url{https://pydantic.dev/pydantic-ai}}, allowing the definition of the tools and orchestrator. The model \texttt{google/gemma-4-26B-A4B-it} used by multiple components is served with the vLLM library\footnote{\url{https://vllm.ai/}}. For the retrieval, we use the Pyserini~\cite{Lin:2021:SIGIR} and PyTerrier\_t5~\footnote{\url{https://github.com/terrierteam/pyterrier_t5}} libraries.
For the clarification need prediction, we use the resources released by the authors of the MuSIc and Zef-CNP models to perform training and inference.\footnote{\url{https://github.com/ChuanMeng/SIP} and \url{https://github.com/lulili0963/Zef-CNP} for MuSIc and Zef-CNP, respectively.}

\section{Submitted Runs}
\label{sec:runs}

We submitted two runs which only differ in the model used for clarification need prediction. 
\begin{enumerate}
    \item \textbf{official\_run\_bert\_llm} uses a BERT model trained for clarification need prediction using the Zef-CNP dataset~\citep{Lu:2025:ECIR}.
    \item \textbf{official\_run\_music\_llm} uses a MuSIc model trained for clarification need prediction using MSDialog~\citep{Qu:2018:SIGIR} and fine-tuned on ClariQ~\citep{Aliannejadi:2021:EMNLP}.
\end{enumerate}

\section{Results}
\label{sec:results}

This section presents the evaluation results of our submitted runs for the iKAT shared task, based on the official task metrics. The evaluation is divided into two parts: system-user interaction quality (Tables~\ref{tab:results_rubric} and \ref{tab:results_interaction}) and passage retrieval performance (Table~\ref{tab:results_passage_ranking}).

\subsection{Dialogue Evaluation}

\begin{table}
    \centering
    \caption{Average rubric level evaluation scores for each user type and overall. The prefix \texttt{official\_} is omitted from run names.}
    \label{tab:results_rubric}
    \begin{tabular}{llccc}
        \toprule
        \textbf{Run} & \textbf{User Type} & \textbf{Engagement} & \textbf{Relevance} & \textbf{Overall Quality} \\ \midrule
        \multirow{5}{*}{run\_bert\_llm} & u1 & 2.75 \scriptsize{$\pm$ 0.80} & 3.47 \scriptsize{$\pm$ 1.45} & 3.16 \scriptsize{$\pm$ 1.32} \\
        & u2 & 3.09 \scriptsize{$\pm$ 0.82} & 2.94 \scriptsize{$\pm$ 1.26} & 2.78 \scriptsize{$\pm$ 1.12} \\
        & u3 & 3.12 \scriptsize{$\pm$ 0.66} & 3.68 \scriptsize{$\pm$ 1.28} & 3.27 \scriptsize{$\pm$ 1.09} \\
        & u4 & 2.91 \scriptsize{$\pm$ 0.89} & 3.53 \scriptsize{$\pm$ 1.50} & 3.14 \scriptsize{$\pm$ 1.25} \\ \cline{2-5}
        & all & 2.98 \scriptsize{$\pm$ 0.81} & 3.40 \scriptsize{$\pm$ 1.40} & 3.08 \scriptsize{$\pm$ 1.20} \\ \midrule
        \multirow{5}{*}{run\_music\_llm} & u1 & 3.24 \scriptsize{$\pm$ 0.81} & 2.60 \scriptsize{$\pm$ 1.22} & 2.44 \scriptsize{$\pm$ 1.04} \\
        & u2 & 3.26 \scriptsize{$\pm$ 0.84} & 2.50 \scriptsize{$\pm$ 1.08} & 2.46 \scriptsize{$\pm$ 0.95} \\
        & u3 & 3.40 \scriptsize{$\pm$ 0.63} & 3.20 \scriptsize{$\pm$ 1.24} & 2.95 \scriptsize{$\pm$ 1.06} \\
        & u4 & 3.28 \scriptsize{$\pm$ 0.77} & 2.69 \scriptsize{$\pm$ 1.15} & 2.53 \scriptsize{$\pm$ 0.94} \\ \cline{2-5}
        & all & 3.29 \scriptsize{$\pm$ 0.77} & 2.74 \scriptsize{$\pm$ 1.19} & 2.59 \scriptsize{$\pm$ 1.01} \\ \bottomrule
    \end{tabular}
\end{table}

\begin{table}
    \centering
    \caption{Average dialogue level evaluation scores for each user type and overall. Criteria: mixed-initiative strategies (Mix), personalization (Pers), information flow (Flow), trustworthiness (Trust), and user satisfaction (Sat). The prefix \texttt{official\_} is omitted from run names.}
    \label{tab:results_interaction}
    \begin{tabular}{llccccc}
        \toprule
        \textbf{Run} & \textbf{\makecell[l]{User\\Type}} & \textbf{Mix} & \textbf{Pers} & \textbf{Flow} & \textbf{Trust} & \textbf{Sat} \\ \midrule
        \multirow{5}{*}{run\_bert\_llm} & u1 & 2.22 \scriptsize{$\pm$ 0.97} & 3.78 \scriptsize{$\pm$ 0.44} & 3.00 \scriptsize{$\pm$ 0.71} & 2.44 \scriptsize{$\pm$ 0.88} & 2.56 \scriptsize{$\pm$ 1.01} \\
        & u2 & 3.30 \scriptsize{$\pm$ 0.67} & 3.80 \scriptsize{$\pm$ 0.42} & 3.00 \scriptsize{$\pm$ 0.47} & 3.20 \scriptsize{$\pm$ 0.42} & 3.00 \scriptsize{$\pm$ 0.67} \\
        & u3 & 3.10 \scriptsize{$\pm$ 0.74} & 4.00 \scriptsize{$\pm$ 0.00} & 3.70 \scriptsize{$\pm$ 0.48} & 3.50 \scriptsize{$\pm$ 0.53} & 3.60 \scriptsize{$\pm$ 0.70} \\
        & u4 & 3.20 \scriptsize{$\pm$ 0.68} & 4.00 \scriptsize{$\pm$ 0.00} & 3.27 \scriptsize{$\pm$ 0.80} & 2.93 \scriptsize{$\pm$ 0.88} & 3.33 \scriptsize{$\pm$ 0.90} \\ \cline{2-7}
        & all & 3.00 \scriptsize{$\pm$ 0.84} & 3.91 \scriptsize{$\pm$ 0.29} & 3.25 \scriptsize{$\pm$ 0.69} & 3.02 \scriptsize{$\pm$ 0.79} & 3.16 \scriptsize{$\pm$ 0.89} \\ \midrule
        \multirow{5}{*}{run\_music\_llm} & u1 & 3.11 \scriptsize{$\pm$ 0.78} & 3.78 \scriptsize{$\pm$ 0.44} & 2.56 \scriptsize{$\pm$ 0.53} & 2.67 \scriptsize{$\pm$ 0.71} & 2.67 \scriptsize{$\pm$ 0.87} \\
        & u2 & 3.30 \scriptsize{$\pm$ 0.67} & 3.80 \scriptsize{$\pm$ 0.42} & 2.80 \scriptsize{$\pm$ 0.63} & 3.20 \scriptsize{$\pm$ 0.42} & 3.00 \scriptsize{$\pm$ 0.82} \\
        & u3 & 3.70 \scriptsize{$\pm$ 0.48} & 3.90 \scriptsize{$\pm$ 0.32} & 3.40 \scriptsize{$\pm$ 0.70} & 3.60 \scriptsize{$\pm$ 0.70} & 3.40 \scriptsize{$\pm$ 0.84} \\
        & u4 & 2.73 \scriptsize{$\pm$ 0.96} & 3.47 \scriptsize{$\pm$ 0.52} & 2.60 \scriptsize{$\pm$ 0.74} & 2.67 \scriptsize{$\pm$ 0.62} & 2.47 \scriptsize{$\pm$ 0.99} \\ \cline{2-7}
        & all & 3.16 \scriptsize{$\pm$ 0.83} & 3.70 \scriptsize{$\pm$ 0.46} & 2.82 \scriptsize{$\pm$ 0.72} & 3.00 \scriptsize{$\pm$ 0.72} & 2.84 \scriptsize{$\pm$ 0.94} \\ \bottomrule
    \end{tabular}
\end{table}

Dialogue evaluation is performed at both the rubric (Table~\ref{tab:results_rubric}) and dialogue (Table~\ref{tab:results_interaction}) levels using an LLM-as-a-judge approach. 

The rubric level assessment evaluates 5 rubrics per conversation with regard to engagement, relevance and usefulness, and overall quality of the system's responses. Results indicate that the run \texttt{official\_run\_bert\_llm} performs better than \texttt{official\_run\_music\_llm} for relevance and overall quality, while the opposite is observed for engagement. 

For the dialogue level assessment, 5 criteria are evaluated per conversation: (1) mixed-initiative strategies, (2) personalization, (3) information flow, (4) trustworthiness, and (5) user satisfaction. 
Focusing on the mixed-initiative strategies criterion, we observe that the main difference occurs for user type u1, where the run \texttt{official\_run\_music\_llm} performs better than \texttt{official\_run\_bert\_llm}. More generally, we note that the average scores for this criterion are around 3 out of 5, indicating a moderate performance of the system with regard to mixed-initiative strategies. Regarding the other criteria, we observe that the run \texttt{official\_run\_bert\_llm} performs better than \texttt{official\_run\_music\_llm} for personalization, information flow, and user satisfaction, with similar performance for trustworthiness. 

Overall, performance differences between the two runs are marginal across all criteria, which does not allow us to conclude which clarification need prediction model is better. Furthermore, a comparison with the other participants' runs remains to be performed to assess the relative performance of our system and the impact of our clarification need prediction models.

\subsection{Passage Ranking Evaluation}

\begin{table}
    \centering
    \caption{Passage ranking results. The prefix \texttt{official\_} is omitted from run names.}
    \label{tab:results_passage_ranking}
    \begin{tabular}{lcccccc}
        \toprule
        \textbf{Run} & \textbf{\makecell[l]{User\\Type}} & \textbf{AP} & \textbf{R@10} & \textbf{P@10} & \textbf{nDCG} & \textbf{nDCG@5} \\ \midrule
        \multirow{4}{*}{run\_bert\_llm} & u1 & 0.0214 &  0.0027 & 0.0424 & 0.1350 & 0.0299 \\
        & u2 & 0.0201 & 0.0015 & 0.0272 & 0.1188 & 0.0343 \\
        & u3 & 0.0183 & 0.0018 & 0.0272 & 0.1034 & 0.0204 \\
        & u4 &  0.0308 & 0.0023 & 0.0370 & 0.1793 & 0.0289 \\ \midrule
        \multirow{4}{*}{run\_music\_llm} & u1 & 0.0108 & 0.0008 & 0.0174 & 0.0617 & 0.0139 \\
        & u2 & 0.0086 & 0.0009 & 0.0152 & 0.0487 & 0.0147 \\
        & u3 & 0.0074 & 0.0010 & 0.0109 & 0.0476 & 0.0061 \\
        & u4 & 0.0194 & 0.0013 & 0.0196 & 0.1019 & 0.0162 \\ \bottomrule
    \end{tabular}
\end{table}

Table~\ref{tab:results_passage_ranking} presents the passage ranking results provided by the organizers. The performance is extremely low across all metrics. 

We hypothesize that this performance is due to the low retrieval rate and the simplicity of the retrieval-reranking pipeline (i.e., BM25 followed by monoT5 re-ranker).
Specifically, the conversational agent provided passages for only $41.5\%$ of user utterances in \texttt{official\_run\_bert\_llm} and $14.5\%$ in \texttt{official\_run\_music\_llm}. This was caused either by the orchestrator bypassing the retrieval tool entirely or experiencing execution failures while producing the final response ($4\%$ and $2\%$ of instances, respectively).

These results indicate that our naive implementation of the orchestrator requires a stricter execution flow to ensure that the retrieval tool is called for user utterances that require a response rather than a clarification question.

\section{Conclusion}
\label{sec:conclusion}

In iKAT 2026, we explored the mixed-initiative ability of a conversational search system, focusing on clarification need prediction. We implemented a naive agentic architecture where the orchestrator agent decides which component to call based on the current state of the conversation, where the tools correspond to the components of a traditional pipeline-based architecture. We submitted two runs that only differ in the model used for clarification need prediction. The results indicate marginal differences between the two runs, although the mixed-initiative performance is moderate for the different user profiles. A comparison with the other participants' runs and human evaluation remains to be performed to assess the relative performance of our system with regards to the different criteria. 

We identify several areas for future experimentation and improvements. Regarding future experiments, having a baseline run without any clarification need prediction model would help to assess whether such models are beneficial for the mixed-initiative ability of the system. Another interesting experiment relates to the clarification question generation, where using a smaller and specifically trained model could be compared to the zero-shot generation with a large model to consider lower resource scenarios. In terms of improvements, the retrieval capability of our conversational system appears as a critical area. Another point concerns repair strategies, we implemented a simplistic one consisting of acknowledging that a response is not possible and asking the user to rephrase their utterance. More sophisticated repair strategies could be implemented to provide more informative responses to the user based on the specific problem encountered.

\section*{Acknowledgments}

We thank Timo Breuer for his helpful comments during the preparation of our submission. This work was supported by Deutsche Forschungsgemeinschaft (407518790) and the German Federal Ministry of Education and Research (BMBF) and Joint Science Conference (GWK) in the PLan\_CV project (03FHP109).

\renewcommand*{\bibfont}{\scriptsize}
\bibliographystyle{abbrvnat}
\bibliography{scai2026-ikat.bib}

\end{document}